# Exploring the Dynamics of Three-Dimensional Lattice Gauge Theories by External Fields

P. Cea* [a] and L. Cosmai [b]

[a] Dipartimento di Fisica dell'Università di Bari, 70126 Bari, Italy

[b] Istituto Nazionale di Fisica Nucleare, Sezione di Bari, 70126 Bari, Italy

We investigate the dynamics of three-dimensional lattice gauge theories by means of an external abelian magnetic field. For the $SU(2)$ lattice gauge theory we find evidence of the unstable modes.

A useful approach to investigate a given physical system is to submit it to an external perturbation. In this note we consider three-dimensional lattice gauge theories in an external abelian magnetic field. In three dimension $g^2$ has dimension of the mass, so the theory is super-renormalizable. Thus, one may hope that the approach to the continuum will be faster than in four dimensions. Moreover in three-dimensional space-time one can perform Monte Carlo simulations on rather large lattices. Indeed we was able to do numerical simulation on lattices which range in size up to $40^3$.

For $SU(2)$ we are interested in the so-called unstable modes. Previous studies in four dimensions[1,2] indicated that one needs sizeable lattices to display the unstable modes. Recently[3] A.R. Levi and J. Poloni claim evidence of the unstable modes in four dimensions. These authors suggest that $L \gtrsim 18$ in four dimensions.

The abelian background fields on the lattice can be introduced by means of an external current[4,5]. In the Euclidean continuum the $SU(2)$ background action reads[5]:

$$S_B = -\frac{1}{2} \int d^3x \; F^{\text{ext}}_{\mu\nu}(x) \left[\partial_\mu A^3_\nu(x) - \partial_\nu A^3_\mu(x)\right] . \quad (1)$$

To discretize the background action we must define the Abelian-like field strength tensor $F^A_{\mu\nu} = \left[\partial_\mu A^3_\nu(x) - \partial_\nu A^3_\mu(x)\right]$ on the lattice. We used the abelian projection[5].

Taking into account the periodic boundary conditions we write

$$F^{\text{ext}}_{\mu\nu}(x) = \sqrt{\beta} \; \sin \theta^{\text{ext}}_{\mu\nu}(x) , \quad (2)$$

$$\theta^{\text{ext}}_{\mu\nu}(x) = \frac{2\pi}{L^2} n(x) , \quad (3)$$

where $n$ is a positive integer. The total action for external constant magnetic field is

$$S = S_W - \beta \sum_x \sin \theta^{\text{ext}}_{12} \; \sin \theta^A_{12}(x) , \quad (4)$$

where $\theta^A_{12}$ is the abelian projected plaquette angle, and $S_W$ is the standard Wilson action. Recently H.D. Trottier and R.M. Woloshyn proposed a different discretization for the abelian field strength tensor[6] (for a comparison with these authors see Ref.[7]).

We are interested in the vacuum energy density. For $SU(2)$ we have

$$E\left(F^{\text{ext}}_{12}\right) = \beta \left[P_s\left(F^{\text{ext}}_{12}\right) - P_t\left(F^{\text{ext}}_{12}\right)\right] \quad (5)$$

where

$$P_s\left(F^{\text{ext}}_{12}\right) = 1 - \frac{1}{2} \operatorname{tr} U_{12}\left(F^{\text{ext}}_{12}\right) \quad (6)$$

$$P_t\left(F^{\text{ext}}_{12}\right) = \sum_{i=1,2} \left[1 - \frac{1}{2} \operatorname{tr} U_{3i}\left(F^{\text{ext}}_{12}\right)\right] . \quad (7)$$

For $U(1)$: $\frac{1}{2}\operatorname{tr} \longrightarrow \operatorname{Re}$. The energy difference is

$$\Delta E\left(F^{\text{ext}}_{12}\right) \equiv E\left(F^{\text{ext}}_{12}\right) - E(0) . \quad (8)$$

---

*Talk presented by P. Cea

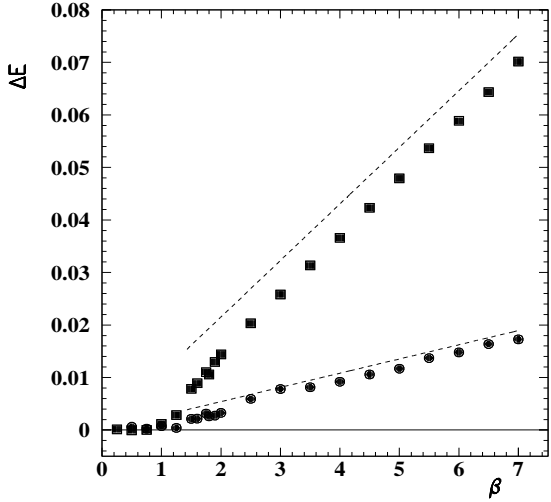

Figure 1. The $U(1)$ vacuum energy density difference versus $\beta$. The lattice size is $L = 16$. Circles and squares correspond to $n = 3$ and $n = 6$ respectively. The dashed line is the classical magnetic energy $\frac{1}{2}(F_{12}^{\text{ext}})^2$.

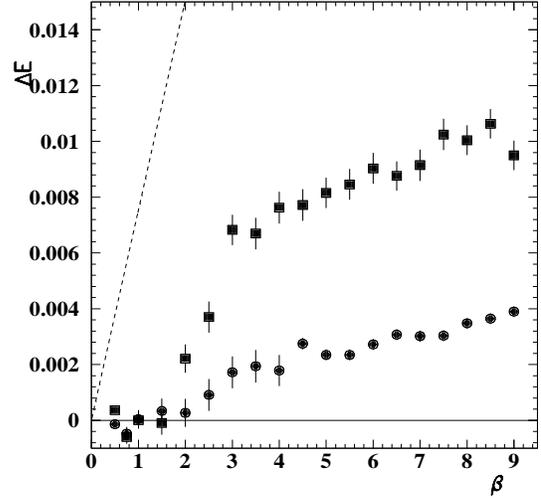

Figure 2. The $SU(2)$ vacuum energy density difference versus $\beta$. The lattice size is $L = 16$. Circles and squares correspond to $n = 5$ and $n = 10$ respectively. The dashed line is the classical magnetic energy for $n = 5$.

In the weak field strength region (which is the region relevant for the continuum limit) we can neglect the electric contribution to the energy density. This allows us to measure directly during Monte Carlo runs the energy difference[8]. We have checked that the two methods give consistent results in the whole range of the applied field strengths. In Figure 1 we dispay the $U(1)$ vacuum energy difference versus $\beta$ for two different values of the external magnetic field. In the strong coupling region we see that the external magnetic field is screened. By increasing $\beta$ the magnetic field penetrates into the lattice. The dashed line is the classical magnetic energy. The agreement with the continuum expectations is rather good for weak field strengths. In Figure 2 we show the $SU(2)$ vacuum energy density versus $\beta$. At first sigth Fig. 2 looks similar to the previous one.

Indeed at strong coupling the abelian magnetic field is screened, and by increasing $\beta$ it penetrates into the lattice. However the vacuum energy is smaller by more than one order of magnitude with respect to the classical magnetic energy (dashed line). This drastic reduction of the vacuum energy can be ascribed to the unstable modes. In Ref.[8] we showed that in the one-loop approximation the vacuum energy density is given by

$$\frac{\Delta E(F_{12}^{\text{ext}})}{g^6} = f(x) \quad , \quad x = \frac{F_{12}^{\text{ext}}}{g^3} \qquad (9)$$

where

$$f(x) = a_{3/2} x^{3/2} + a_2 x^2 \quad . \qquad (10)$$

Neglecting the unstable modes Eq.(9) coincides with the one-loop effective potential:

$$a_{3/2} = \frac{1}{2\pi}\left[1 - \frac{\sqrt{2}-1}{4\pi}\zeta(3/2)\right] \;,\; a_2 = \frac{1}{2} \;. \quad (11)$$

On the other hand, after stabilization of the unstable modes we get

$$a_{3/2} = \frac{\sqrt{2}-1}{8\pi^2}\zeta(3/2) \;, \qquad a_2 = 0 \;. \quad (12)$$

Moreover the chromomagnetic field is strongly screened by the unstable modes. Indeed we found[8]

$$\frac{F_{12}^3}{F_{12}^{\text{ext}}} = \frac{1}{2\pi}\left(\frac{F_{12}^{\text{ext}}}{g^3}\right)^{-\frac{1}{2}} \;. \quad (13)$$

Equation (13) is to be compared with the result we would have obtained without the unstable modes

$$\frac{F_{12}^3}{F_{12}^{\text{ext}}} = 1 \;. \quad (14)$$

In Figure 3 we show the scaled energy density versus the scaling variable $x$. We see that the scaling law Eq.(9) is satisfied quite well by the data. Moreover the data are in satisfying agreement with Eq.(12) (dashed line) in the weak field strength region, but are in striking disagreement with the one-loop effective potential (dotted line). We have also measured the chromomagnetic field strength (see Fig. 3 of Ref.[8]). Again we found that the data were close to Eq.(13). The disagreement with our theoretical calculations is due to a very small ($a_2 \simeq 0.015$) classical-like term in the energy density. In view of the fact that the absence of the classical term in the energy density can be obtained only in the thermodynamical limit[8] we feel that our Monte Carlo data support the evidence of the unstable modes. In particular, we showed that the states with a constant chromomagnetic background field are not energetically favoured with respect to the perturbative ground state. Thus these states are not relevant for the dielectric model of confinement. However, the drastic reduction of the energy due to the unstable mode condensation leaves open the possibility of a confining vacuum which behaves like a ferromagnet with random oriented domains.

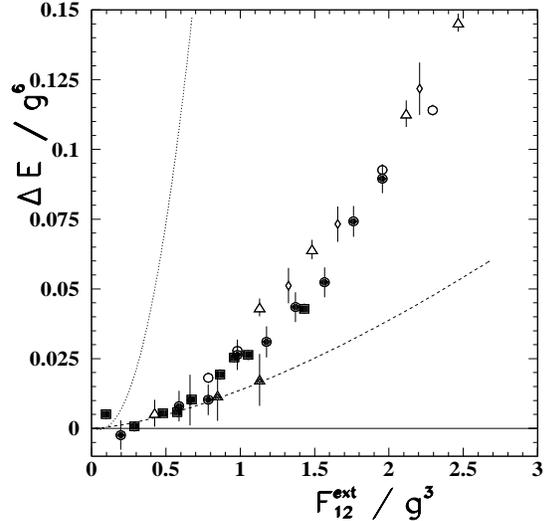

Figure 3. The scaled vacuum energy density versus $x$. Full and open symbols refer to $L = 20$ and $L = 40$ respectively. Squares correspond to $\beta = 7$, circles to $\beta = 10$, triangles to $\beta = 12$, and diamonds to $\beta = 15$. The curves are discussed in the text.